# J-PARC MUSE H-line optimization for the g-2 and MuHFS experiments

A Toyoda[1,4], Y Fujiwara[3], Y Fukao[1], O Kamigaito[3], N Kawamura[1], Y Matsuda[2], T Mibe[1], T Ogitsu[1], N Saito[1], K Sasaki[1], K Shimomura[1], M Sugano[1], K Tanaka[2], D Tomono[3], and H A Torii[2]

[1]KEK, 1-1, Oho, Tsukuba, Ibaraki, 305-0801, JAPAN
[2]The University of Tokyo, Institute of Physics, Tokyo, JAPAN
[3]RIKEN, Saitama, JAPAN

**Abstract.** Significant deviation of the anomalous magnetic moment value (g-2) observed by the muon g-2 experiment should be confirmed by the other experiment. This value is experimentally determined by frequency difference observed by the g-2/EDM experiment and muon magnetic moment observed by the muonium hyperfine splitting experiment (MuHFS). Both two experiments are planned to be performed at H-line of the J-PARC/MUSE under construction. We optimized the beamline layout for each experiment with G4beamline. For both experiments, statistics is the most important, thus beamline transmission efficiency should be maximized. Especially for the g-2, the purpose of the present effort is to compromise between small beam size and small leakage field. For the MuHFS, it is crucial to minimize leakage field at around final focus position, and to get all stopped muons within good field region of MuHFS magnet. Conceptual design of the several final focusing systems will be presented. *contribution ( or invited paper) to NUFACT 11, XIIIth International Workshop on Neutrino Factories, Super beams and Beta beams, 1-6 August 2011, CERN and University of Geneva*
(Submitted to IOP conference series)

## 1. Introduction

The 3.4 sigma deviation of anomalous muonic magnetic moment (g-2) from the standard model (SM) prediction reported by the BNL E821 experiment is expected to be an indication of new physics, and should be confirmed by the other experiment with higher accuracy. Also it is needed to increase experimental accuracy of ratio of muonic magnetic moment to proton magnetic moment ($\mu_\mu/\mu_p$) to achieve the g-2 precision improvement. This value is determined by the muonium hyperfine splitting measurement (MuHFS). The other topic is that the 3.9 sigma deviation of the positronium hyperfine splitting (PsHFS) measurement from the SM prediction also indicates new physics. Accuracy of the MuHFS is about 40 times worse than that of PsHFS, and thus this should be increased more.

Under the circumstances, two new experiments, g-2/EDM experiment [1] to measure g-2 and MuHFS experiment [2] to measure MuHFS, are proposed and planned to be performed at H line of the J-PARC/MUSE facility as shown in Figure 1. Basic strategy and optimization of the H line is reported in reference [3]. This article mainly focused on the final focusing design of these two experiments.

---

[4]  E-mail: akihisa.toyoda@j-parc.jp

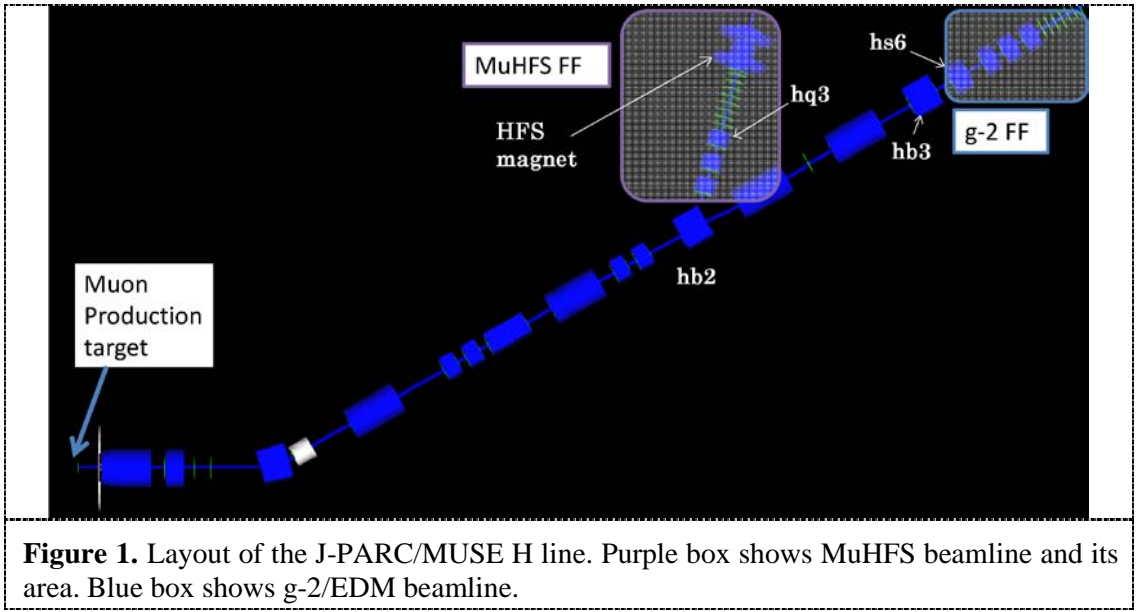

**Figure 1.** Layout of the J-PARC/MUSE H line. Purple box shows MuHFS beamline and its area. Blue box shows g-2/EDM beamline.

## 2. g-2/EDM beamline optimization

In this section we discuss the g-2/EDM beamline optimization done by using G4BEAMLINE program [4]. For this experiment, the following conditions are important to be satisfied.

- *Transmission efficiency should be as high as possible.* Statistics is the most important issue of this experiment.
- *Focus point size should be less than 4 cm in diameter.* This experiment installs muonium production target at the beamline final focus point. The muonium emitted from this target is irradiated with a laser in order to produce the ultra slow muon beam. Small beam spot size helps to increase the slow muon intensity.
- *Leakage field at final focus position should be as small as possible.* The slow muon has low momentum at around the final focus, thus the leakage field disturbs the slow muon trajectory.

The other parameters such as x/y beam angle distribution are not so much important. Possible final focus systems are evaluated in the following subsection.

### 2.1. Final focus system with 1 solenoid case

The simplest way to make small beam spot size is to use solenoid magnet. A solenoid magnet of hs6 is installed just downstream of the last bending magnet of hb3. Transmission efficiency is as high as 87.5 % inside 40 $\phi$ at final focus position. Focus point size is small 1.7 cm $\sigma$ for x, 1.8 cm $\sigma$ for y. However, leakage field of hs6 magnet is the problem. Figure 2 shows the magnetic field along beam axis. To make small beam spot, maximum field of hs6 is as high as 0.42 T. This leads to short focal length of 0.4 m, thus leakage field at the final focus point is as high as 0.07 T. To use this option, we need to design magnetic shield for the hs6 magnet carefully.

### 2.2. Final focus system with 3 quadrupole magnets

To solve the leakage field problem mentioned in the above subsection, we have an option to use 3 quadrupole magnets as the focusing system. Last component is quadrupole, thus the leakage field along the beam axis is zero. The maximum field of the quadrupole magnet is lower than that of the solenoid magnet therefore the leakage field at around the beam axis is also expected to be smaller.
The result of the optimization is that the transmission efficiency is 63.7 %, and the beam size is 3.5 cm $\sigma$ for x, 1.8 cm $\sigma$ for y. This result is not good enough therefore we cannot use this option.

### 2.3. Final focus system with 1 solenoid and 3 quadrupole magnets

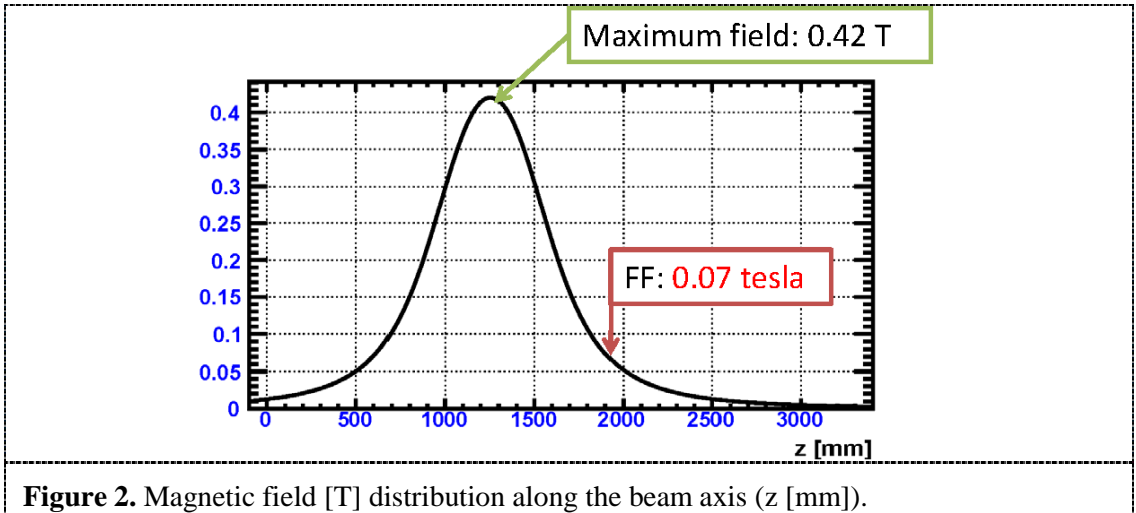
**Figure 2.** Magnetic field [T] distribution along the beam axis (z [mm]).

The smallness of the transmission efficiency of the above 3 quadrupole system comes from its small focusing power. The solenoid system has the large leakage field. To solve these problems, we devise the following method as shown in Figure 3. First of all, we make a small focus spot by using the strong focusing power of an hs6 solenoid magnet. This small focus spot is transported to the final focus by using 3 quadrupole magnets. This method has an advantage of a small focus spot and a small leakage field. The estimated transmission efficiency is as high as 83.4 %, and the final focus size is as small as 2.1 cm σ for x, 1.1 cm σ for y. This method will be applicable to the g-2/EDM experiment.

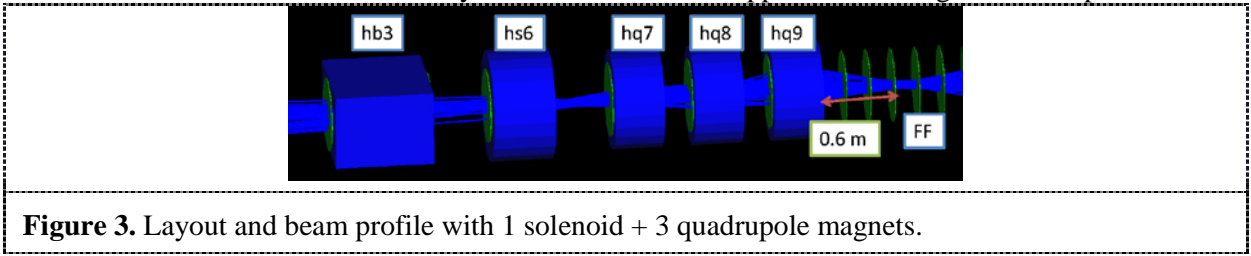
**Figure 3.** Layout and beam profile with 1 solenoid + 3 quadrupole magnets.

**3. MuHFS beamline optimization**
In this section we discuss the MuHFS beamline optimization. For this experiment, the following conditions are important to be satisfied.
- *Transmission efficiency should be as high as possible*. Statistics is the most important also for this experiment.
- *Focus point size should be less than 6 cm in diameter*. For this experiment, the beam focusing is not so much serious issue compared to the above g-2/EDM experiment, because we installed the HFS solenoid magnet and the strong magnetic field as high as 1.7 T eventually focuses the surface muon beam. This condition is decided to make sure that all muons stop inside good field region of the HFS magnet (20 cm in diameter).
- *Leakage field at final focus position should be less than 1.7 gauss*. One of the key issues of this experiment is to make a large and homogenious magnetic field region. This limit of 1.7 gauss comes from the limitation of the magnetic field correction power by the shim coil [5].
- *Muon stopping distribution should be inside good field region of the HFS magnet (30 cm long)*. This condition will limit the muon momentum spread.

The other parameters such as x/y beam angle distribution are not so much important. Possible final focus systems are evaluated as follows.
The most effective way to suppress the leakage field of the beamline magnet is to keep the focal length as long as possible. To achieve this long focusing, we devise the following method. At first, we prepare almost parallel beam by using 3 quadrupole magnets as shown in Figure 4.

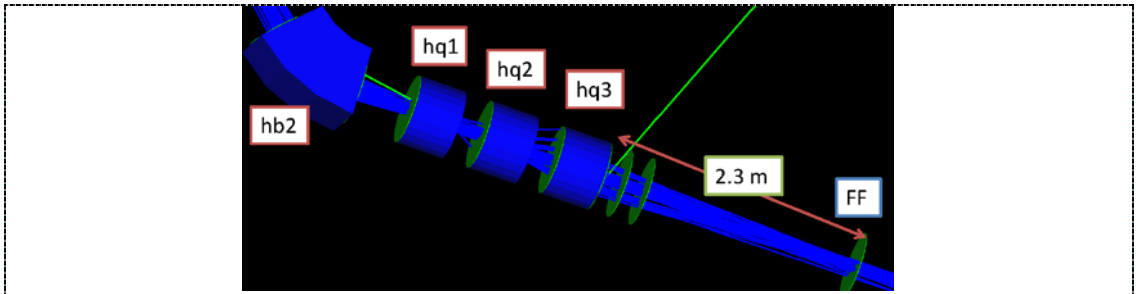
**Figure 4.** Layout and beam profile of the MuHFS final focus beamline.

Next we put the HFS magnet at the final focus point. The created magnetic-field distribution is consistent with Opera calculation. The HFS magnet design is described in detail in reference [5]. One of the possible problems is effect of the leakage field of the strong HFS magnetic field to the muon beam transport. This leakage field is carefully investigated as shown in Figure 5, and the leakage field at a position of the last quadrupole magnet (hq3) is estimated to be as small as 0.005 T. We also checked the MuHFS leakage field effect to the beam profile at the hq3 end. It is confirmed that the MuHFS leakage field does not affect the beam transport seriously. Finally, we estimated the transmission efficiency and the beam profile at the final focus point with the MuHFS magnet. The result is that the transmission efficiency is as high as 93.6 % and final focus spot size is as small as 1.3 cm $\sigma$ for x, 1.3 cm $\sigma$ for y. It is concluded that this method will be applicable to the MuHFS final focus system.

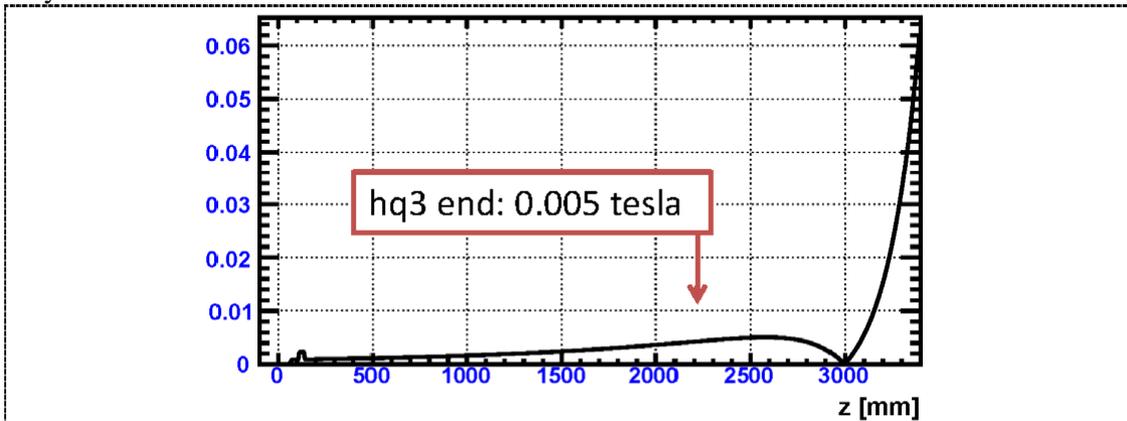
**Figure 5.** Magnetic field [T] distribution along the beam axis (z [mm]).

### 4. Summary and future plan

For the g-2/EDM beamline, the final focusing with 1 solenoid and 3 quadrupole magnets is effective, the transmission of 87.5 % is high and the beam size of 2.1 cm $\sigma$ for x, 1.1 cm $\sigma$ for y is small enough. For the MuHFS beamline, the final focusing with 3 quadrupole magnets works well, the transmission is as high as 93.6 % and the beam size is as small as 1.3 cm $\sigma$ for x, 1.3 cm $\sigma$ for y.

In the future, we need to estimate leakage field from the quadrupole magnets at a final focus point for both beamlines. For the MuHFS beamline, it is necessary to evaluate the muon stopping distribution.